# Learning Atoms from Crystal Structure

*Andrij Vasylenko[a], Dmytro Antypov[a], Sven Schewe[b], Luke M. Daniels[a], John B. Claridge[a], Matthew S. Dyer[a], Matthew J. Rosseinsky\*[a]*

[a]     Department of Chemistry
University of Liverpool
Crown Street, L69 7ZD, United Kingdom
E-mail: rossein@liverpool.ac.uk

[b]     Department of Computer Science
University of Liverpool
Ashton Building, L69 3DR, United Kingdom

**ABSTRACT:** Computational modelling of materials using machine learning (ML) and historical data has become integral to materials research across physical sciences. The accuracy of predictions for material properties using computational modelling is strongly affected by the choice of the numerical representation that describes a material's composition, crystal structure and constituent chemical elements. Structure, both extended and local, has a controlling effect on properties, but often only the composition of a candidate material is available. However, existing elemental and compositional descriptors lack direct access to structural insights such as the coordination geometry of an element. In this study, we introduce Local Environment-induced Atomic Features (LEAFs), which incorporate information about the statistically preferred local coordination geometry at an element in a crystal structure into descriptors for chemical elements, enabling the modelling of materials solely as compositions without requiring knowledge of their crystal structure.

In the crystal structure of a material, each atomic site can be quantitatively described by similarity to common local structural motifs; by aggregating these unique features of similarity from the experimentally verified crystal structures of inorganic materials, LEAFs formulate a set of descriptors for chemical elements and compositions. The direct connection of LEAFs to the local coordination geometry enables the analysis of ML model property predictions, linking compositions to the underlying structure-property relationships.

We demonstrate the versatility of LEAFs in structure-informed property predictions for compositions, mapping of chemical space in structural terms, and prioritisation of elemental substitutions. Based on the latter for predicting crystal structures of binary ionic compounds, LEAFs achieve the state-of-the-art accuracy of 86%. These results suggest that the structurally-informed description of chemical elements and compositions developed in this work can effectively guide synthetic efforts in discovering new materials.

## I. INTRODUCTION

The approaches to description of chemical elements, ranging from historical methods like Döbereiner's Triads, Newlands' Octaves, and Mendeleev's periodic table to modern variations of the Pettifor scale, depending on the criteria employed, offer various insights into relationships between elements and their roles in chemical reactions and compound formation[1–6]. More recently, elemental descriptors have evolved into multi-dimensional spaces[7–13], advancing computational modelling of connections between elements, their properties, and the materials they constitute. Detailed numerical descriptions of chemical elements facilitate addressing critical materials science challenges: developing metrics for mapping chemical space[14–21], modelling composition-structure-property relationships[22–31] and materials discovery, including through design by similarity[32–41]. Quantification of similarity between chemical elements arises from elemental descriptors and drives atomic substitution-based design for novel materials at scale[35,40–42]. Incorporating structural insights into representation of chemical elements can significantly enhance the efficiency of materials modelling[39]. Materials structure, defined by i) compositional content, ii) atomic coordination – interatomic distances, iii) atomic positions in a unit cell – angles between central atoms and their coordinations, determines stability and properties of materials. Crystal structure is critical for evaluating novel candidates for materials discovery, yet establishing the relationship between candidate chemical composition and optimal structure is a recognised challenge[43–45]. The state-of-the-art elemental descriptors incorporate various aspects of the material structure, such as compositional content (Atom2Vec), atom connectivity in crystal graph (MEGNet, SkipAtom), and implicit structural information learnt through machine learning of scientific literature (Mat2Vec, MatScholar), however, none of the available elemental descriptors offers direct and explicit access to the geometric aspects of materials structure.

In this study, we explore a novel approach to explicitly incorporate geometrical local structural information for describing chemical elements and materials compositions, resulting in the creation of Local Environment-induced Atomic Features (LEAFs). The LEAFs maintain a direct and explicit relationship between chemical elements and preferred structural characteristics such as atomic coordination and local structure motifs and we demonstrate how this direct connection can help explain machine learning models for materials property predictions. We further employ this link to address other structure-induced challenges in materials science such as derivation of a metric for mapping chemical space in structural terms, and selecting elemental substitutions for novel materials design.

In the LEAFs approach, we hypothesise that atomic properties, and hence their descriptors, can be deduced from the nature of their local structure environments in crystalline inorganic compounds. The determination of these descriptors hinges upon the definition of locality and atomic neighbourhood[46], where each atomic site in a crystal structure of a material can be described in terms of its similarity to the common structural motifs. The atomic site is first described by the interatomic distance-based coordination number (CN) of neighbouring atoms[34]; then for each CN, the geometrical arrangements of the neighbouring atoms can determine similarity to one of the common motifs, e.g., whether a CN2 arrangement is linear or water-like, a CN4 arrangement is tetrahedral or square planar, etc., up to CN12 (Figures S1, S2). Quantification of the similarity between the local structure motifs is performed by comparing interior and dihedral angles for each atomic site in a local structure and the 37 selected common structural motifs presented in Ref.[47], using the angle-based similarity metrics[34,47] (Supplementary Eq. 1, 2). Thus, each atomic site can be represented with a set of 37 numbers, each determining similarity to one of the 37 common motifs; this set of numbers is further used as the atomic site's unique vector identifier. In Figure 1, for the example of a Mg atom in MgO, the local structure environment is compared to the CN6 structural motifs. Concatenation (denoted as ∥ in Fig.1) of the similarity values $s(CN)$ to all common motifs within different CNs produces a Mg-site identifier in MgO – vector $\mathbf{a}$(Mg | MgO); in this particular vector, for all but three coordination environments in CN6, $s = 0$.

Using this approach, we examine the local structure environments for all atomic sites of the 86 most common chemical elements across the experimentally studied materials reported in the Inorganic Crystal Structure Database (ICSD)[48]. For each element, the 37 similarity values were collated for all individual atomic sites containing that element across all considered structures. The mean was then taken for each of the 37 coordination environments, resulting in 37 values which form a vector-descriptor for that element, e.g., $\mathbf{a}(\text{Mg} \mid \text{ICSD})$ in Figure 1. Carrying out this procedure for all 86 elements produces the LEAFs.

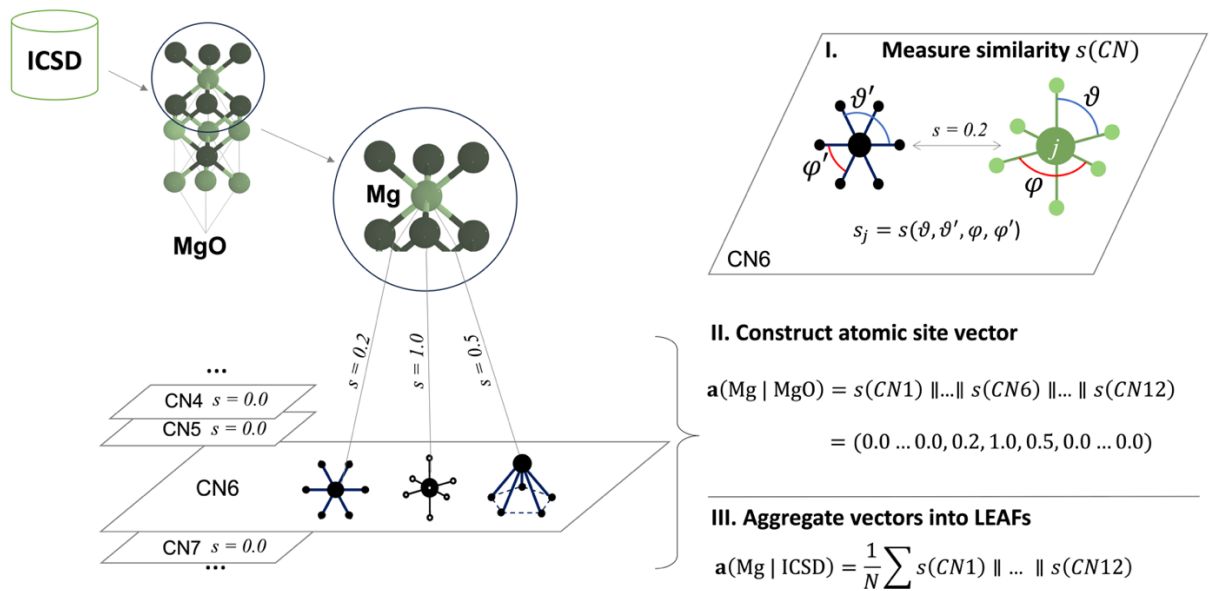

**Figure 1.** Schematic calculation of local environment-induced atomic features (LEAFs). Similarities of the atomic local structure environments in crystal structures are calculated for the common structural motifs[47] within different coordination numbers (CNs), using angle-based similarity metrics[34,42]. For the example of the six-coordinate Mg atom in MgO, similarities, $s$, is zero ($s = 0.0$) to all common motifs, except for the three structural motifs in CN6: hexagonal ($s = 0.2$), octahedral ($s = 1$), and pentagon pyramidal ($s = 0.5$) motifs. Concatenation (symbol $\parallel$) of the similarity values for all structural motifs in all considered CNs produces a local environment vector for an atom in a crystal structure, e.g., for Mg in MgO, $\mathbf{a}(\text{Mg} \mid \text{MgO})$. Collecting these vectors for the 86 most common chemical elements in the crystal structures reported in Inorganic Crystal Structure Database (ICSD)[43], and averaging them over the corresponding occurrences, $N$, of each element produces a set of LEAFs for chemical elements.

## II. SIMILARITY OF CHEMICAL ELEMENTS, COMPOSITIONS AND CRYSTAL STRUCTURES

Numerical representation of chemical elements and compositions determine the quality and efficiency of computational modelling of materials. We study the LEAFs' ability to represent chemical elements in comparison to nine other popular elemental descriptors[7,9–12,33] and include random representation as a baseline. Using these descriptors, chemical elements can be vectorised and compared via cosine similarity (Figure S5). Similarity between elements $X, X'$ can be associated with the degree of probability of elemental substitution in a chemical compound while retaining its crystal structure:

$$p(X, X') = \frac{e^{\cos(X,X')}}{Z} \quad (1)$$

where $Z$ is the partition function; this enables prediction of the crystal structure type based on the probability of elemental substitution and may guide the high-throughput design of materials[5,6,33,38,39,42,49].

We employ the test for predicting crystal structures of binary compounds proposed in Ref.[49] to compare the efficacy of the elemental descriptors. For this test, the set of 494 binary ionic solids reported in Materials Project (MP) is reduced by excluding metallic materials and polymorphs to form a set of 100 AB ionic solids, which match four structure types: CN8 CsCl, CN6 rock-salt, CN4 zinc blende and CN4 wurtzite, using labels from Materials Project[50]. The uneven distribution of structure types in this dataset impedes evaluation of model performance through accuracy in imbalanced classification tasks[51]. To address this, we computed the Matthews' correlation coefficient (MCC)[52] providing a more balanced evaluation of performance for all elemental characteristics studied in [49]. In this task, where for each composition in the test set the structure type is predicted based on the most likely substitution, according to Eq. 1, into the remaining 99 compositions in the test set, LEAFs increase the best values achieved to date (Table 1).

**Table 1.** LEAFs' performance in a multi-class classification task among other elemental features

| Features | Origin of descriptors | Acc.,% | MCC |
|---|---|---|---|
| **LEAFs** | Local coordination geometry in ICSD | **86** | **0.72** |
| MatScholar[9] | ML-derived from literature | 81 | 0.63 |
| Mat2Vec[10] | ML-derived from literature | 80 | 0.60 |
| Atom2Vec[13] | ML-derived from compositional content | 79 | 0.59 |
| GNoME[41] | Prediction of elemental substitution from DFT energies in GNoME | 79 | 0.58 |
| Magpie[7] | Elemental physical characteristics | 78 | 0.54 |
| Oliynyk[8] | Elemental physical characteristics | 75 | 0.50 |
| MEGNet[11] | ML-derived from atom, bond and graph attributes in MP | 73 | 0.45 |
| SkipAtom[12] | ML-derived from atom connectivity graphs in MP | 68 | 0.35 |
| Random | Random numbers | 58 | 0.22 |
| Hautier[33] | Prediction of elemental substitution from DFT energies in MP | 54 | 0.28 |

The enhanced crystal structure classification suggests LEAFs' capability to capture chemical trends. To illustrate this, we plot t-distributed Stochastic Neighbour Embedding (t-SNE) maps of LEAFs representations for chemical elements Figure 2a. Noteworthy trends include clustering of the elements belonging to the same group of the periodic table (colour-coding) or to specific families, such as halogens, chalcogens, metals, metalloids, and noble gases (symbols); the size of the markers corresponds to the atomic number.

In contrast to the random number descriptors (Figure S7), elemental descriptors based on physical and chemical elemental characteristics[7,8], and data-derived vectors[9–12] can effectively organise chemical elements[49], offering insights specific to their properties. In the case of LEAFs, the observed grouping of elements based on their local environments implies similarities in element-specific local structures across experimentally realised inorganic materials: similarity of $3d$ and $4f$ elements, Li, Mg and $3d$ metals, Ca, Y and $4f$ metals, etc. (Figure S5).

These qualitative insights into chemical similarity arising from purely geometrical description of local coordination align with the observations derived with ML of local structural topology[39]. Furthermore, to confirm the LEAFs' ability to recognise chemical patterns beyond elemental grouping, we represent chemical compositions in ICSD as vectors, by summing the weighted elemental LEAFs according to stoichiometry in a chemical formula, e.g., $Li_{0.375}P_{0.125}O_{0.5}$ can be represented with a vector:

$$\mathbf{a}_{Li_{0.375}P_{0.125}O_{0.5}} = 0.375\mathbf{a}_{Li} + 0.125\mathbf{a}_{P} + 0.5\mathbf{a}_{O}, \quad (2)$$

where $\mathbf{a}_X$ is the LEAF vector for the corresponding element X.

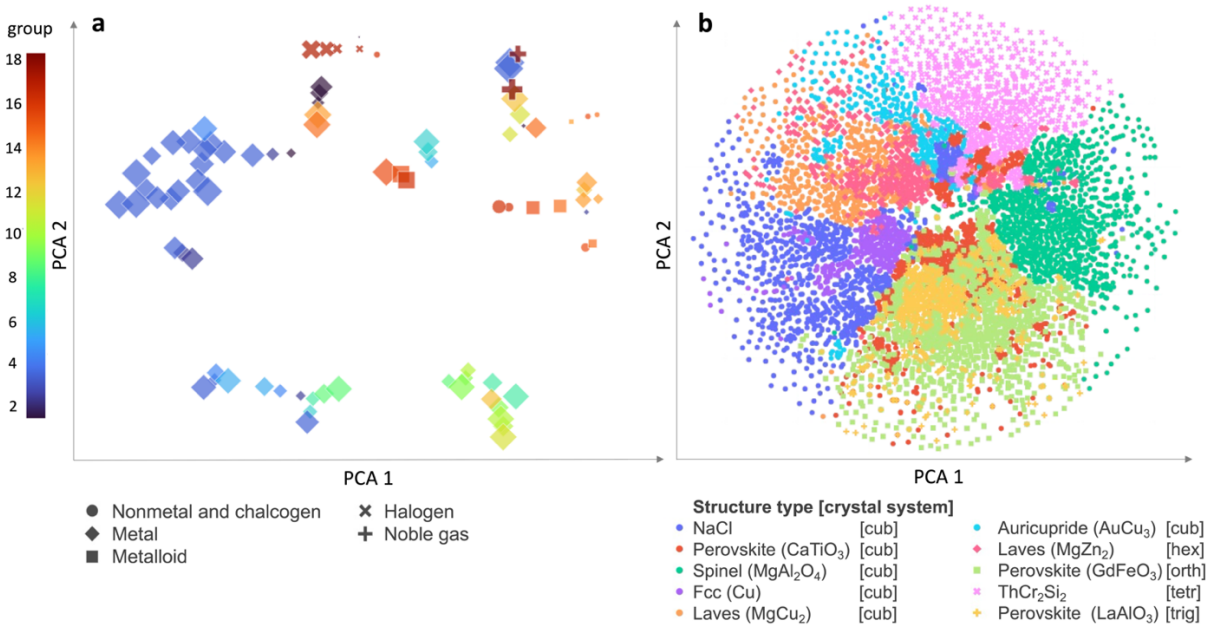

**Figure 2.** The LEAF representation reveals chemical trends for the elements (a) and for compositions in ICSD (b). (a) t-distributed Stochastic Neighbour Embedding (t-SNE) map of elements reveal chemical trends: elements belonging to specific families, such as halogens, chalcogens, metals, metalloids, and noble gases (symbols) and different periodic table groups (colour-coding) cluster together; the marker size denotes atomic number ; (b) Compositions forming the ten most populous structure types in ICSD[49] are represented with LEAFs as in Eq. 2 and plotted in two principal dimensions of the t-SNE map, displaying clustering patterns based on structure type and crystal system: Cu-like structure type within the fcc system (purple circles), the Perovskite (LaAlO$_3$) structure type within the trigonal system (mustard crosses), and the Laves (MgZn$_2$) structure type within the hexagonal system (raspberry diamonds) the ThCr$_2$Si$_2$ structure type in the tetragonal system (pink crosses) and the rock-salt structure type (blue circles) each occupy distinctive areas of the map. The observed patterns suggest that distance in multi-dimensional LEAFs can be used for structural comparison of compositions and design by similarity.

The t-SNE map of the subset of the ten most common structure types in ICSD with compositions represented with LEAFs as in Eq. 2 illustrates the organisational patterns of structure types and crystal systems (Figure 2b). Notably, distinct densely packed clusters representing various structure types are evident: using notations from ICSD, the clusters include the Cu-like structure type within the fcc system (depicted by purple circles), the Perovskite ($LaAlO_3$) structure type within the trigonal system (represented by mustard crosses), and the Laves ($MgZn_2$) structure type within the hexagonal system (depicted by raspberry diamonds). Broader distributions, such as the $ThCr_2Si_2$ ($CeGa_2Al_2$, $BaAl_4$) structure type in the tetragonal system (marked by pink crosses) and the rock-salt structure type (represented by blue circles) are observed, each occupying distinctive areas of the map. Less represented structure types, omitted in Figure 2b for clarity, also demonstrate clustering in analogous t-SNE maps, built with LEAFs (Figure S8). The observed patterns indicate that the multi-dimensional space distance defined by LEAFs, which can be measured, for example, as Euclidean, Wasserstein or other metric distance between compositions represented with LEAFs, can be a metric for structurally-informed comparison between materials defined only by their composition (Eq. S.3), complementing other efforts for effective mapping of chemical space[15–17,19–21,50].

## II. CONNECTING PROPERTIES WITH STRUCTURAL INSIGHTS FOR MATERIALS COMPOSITIONS

LEAFs' potential for representing materials compositions in structural terms can be used in predicting materials properties and uncovering the relationships between the properties and local structure environments in materials, described solely by their compositions. We illustrate this in classification of the Li-ion conducting materials by ionic conductivity, which is reported to strongly depend on the local structural coordination of lithium[51,52]. In a prior study[53], 403 compositions with reported conductivity at room temperature were vectorised via literature-derived elemental descriptors mat2vec[10], and classified into two conductivity classes (below and above $\sigma = 10^{-4}$ S/cm) using a neural-network-based model (CrabNet[54]), achieving an average accuracy of 81% and MCC of 0.47 over 5-fold cross-validation. The underlying elemental descriptors learnt via ML approaches strongly inhibit interpretability of the arising composition-property relationships[55]. In contrast, LEAFs structural insights can emphasise the critical aspects of atomic coordination environments in materials' structures that correlate with their properties, such as ionic conductivity. The importance of various structural aspects can be highlighted through feature selection by using LEAFs with methods such as random forest[56] or neural networks with feature sparsity[57]. The random forest model with LEAFs demonstrates the same average classification accuracy for ionic conductivity of 81% achieved above and MCC of 0.62; by calculating the information entropy gain when selecting different features. In order to highlight the features specific separately to ions of lithium and other species, we expand the compositional representation (2) with concatenation of LEAFs for lithium and a sum of cations, resulting in vectors of double the length, e.g., $Li_7La_3Zr_2O_{12}$ can be represented as

$$\mathbf{a}_{Li_{0.291(6)}La_{0.125}Zr_{0.83(3)}O_{0.5}} \sim 0.292\mathbf{a}_{Li} \parallel (0.125\mathbf{a}_{La}+0.833\mathbf{a}_{Zr}), \qquad (3)$$

where symbol ∥ denotes concatenation, resulting in a 74-bit vector, each elemental vector **a** represented by 37-bit LEAFs. This discriminative power between elemental species comes at a cost of reduced accuracy to 75% (MCC, 0.47), but provides clear indication of contribution of elemental structural motifs to determining ionic conductivity (Figure 3a). Notably, analogous concatenation of LEAFs for lithium and a sum of anions offers the same accuracy of 75% (MCC, 0.47).

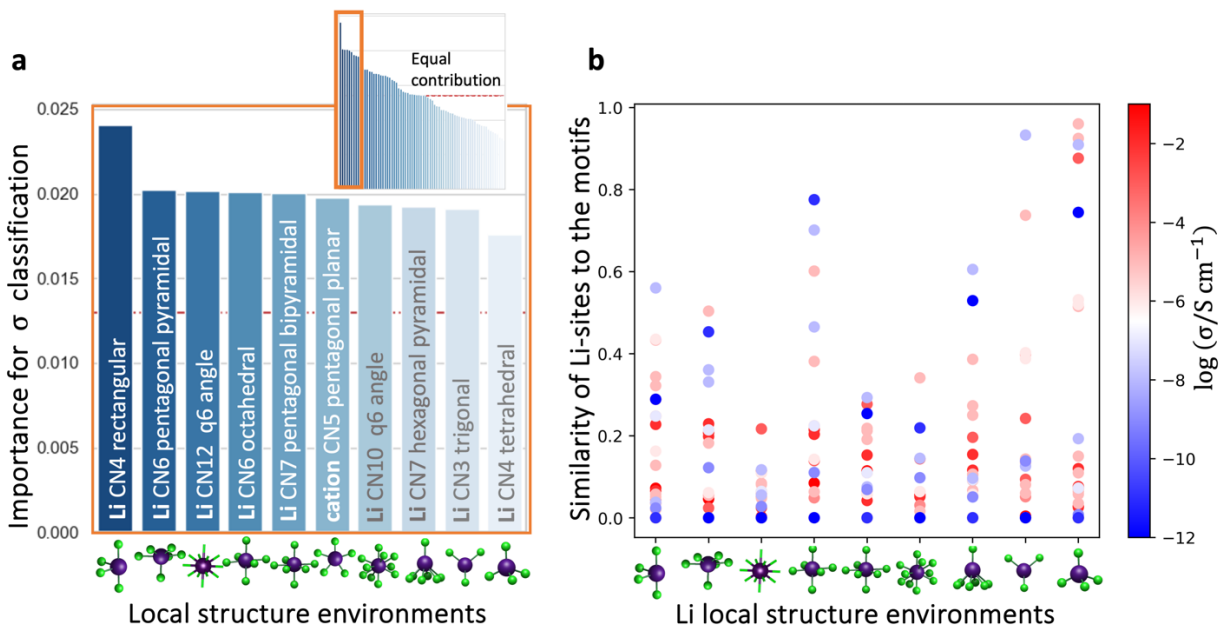

**Figure 3.** Importance of structural environments for classifying materials' ionic conductivity. a Structural insights from LEAFs can highlight the local motifs that influence materials properties predictions: feature importance can be calculated using random forest model in supervised classification, considering conductivity of chemical compositions in Li-ion database[53]. Inset illustrates the contribution of all local structure environments in comparison to equal contribution (dashed line). b In Li-conducting materials, there is a wide distribution of Li local structure environments, demonstrating the absence of a specific preferred Li coordination associated with high Li-ion conductivity.

This may be explained through the feature importance, according to which the majority of top-contributing features are associated with lithium: 29 out of 34 above the equal, uniform contribution line in Figure 3a, and nine out of top ten. This is consistent with 72% accuracy (MCC, 0.42) achieved for classification of Li-ion-conductivity, based on compositional representation solely with lithium content, i.e., $Li_7La_3Zr_2O_{12}$ is represented as $0.292\mathbf{a}_{Li}$, according to fractional Li content. We analyse the crystal structures in the Li-ion conductors database in terms of the similarity of the Li atom sites to the top nine local structure motifs, rendered important for conductivity classification in Figure 3a. The diverse array of Li site local structure environments in Li-conducting materials (Figure 3b, Figure S9) challenges the notion of a specific Li coordination determining Li-ion conductivity, including the widely discussed tetrahedral coordination, as suggested in the literature[51,52]. This observation underscores the significance of considering the collective influence of various local environments of the constituent atoms on materials properties[58].

Furthermore, LEAFs can be integrated with neural network-based models to enable learning the individual sets of elemental descriptors specific to predicting a particular property of materials from the local structure environments. This alignment can be achieved through coupling and end-to-end training of the integrated models[29]. To implement this, we utilise multi-hot encoding to represent the full information regarding elemental local environments across material structures in ICSD in a format easily interpretable by machine learning algorithms. One-hot encoding can represent real values by discretising continuous range values into predefined bins, where only one bin (hot) is set to 1, and the position of this bin indicates the value, for example, numbers 0.0, 0.5 and 1.0 can be represented as strings (1 0 0), (0 1 0) and (0 0 1), respectively, in the 3-bit one-hot encoding scheme. Similarly, we can represent each of the considered common motifs and each individual similarity value, *s*, ranging from 0.001 to 1, with three digits of precision as 1,000-bit vectors. We note that the exact vector length does not appear to have a major effect on the results and re-doing the experiment with 100-bit vectors yielded similar results. In the considered example of the MgO crystal (Figures 1, 4), the similarities of Mg in the octahedral environment to the CN6 motifs, *s* = 0.2, 0.5, 1, can be represented as 1,000 bit binary vectors with 1s in positions 200, 500, and 1000, respectively; for the other 34 motifs, the Mg atom in MgO has similarity *s* = 0, and hence the corresponding binary vectors will have 1s in the first positions. Concatenating these binary vectors for all 37 motifs results in a sparse 37,000-bit multi-hot vector with exactly 37 1s in the corresponding positions, encoding the similarity values of Mg in MgO. This representation also affords encoding of those materials where an atom is found in more than one coordination environment; such materials are represented with binary vectors with more than one bit set to 1. We then use the binary vector to collect all occurring similarity values for Mg local environments in all Mg-containing materials reported in ICSD and populate the bins in the corresponding positions with 1s. Doing this for all chemical elements, we encode each element as a 37,000-bit binary string, where 0s denote the absence and 1s the presence of a similarity value to one of the motifs within the corresponding local environment in ICSD. We illustrate this matrix of local elemental environments conceptually by black and white pixels, representing ones and zeros, respectively for the subsets of elements and their similarities to local environments in Figure 4 (d), and in Figures S1-S3, where more detail is given. The full matrix of local elemental environments is 37,000 columns of binary strings by 86 rows of considered elements. This matrix is then pruned to remove all-zero columns and used as a source for nonlinear learning of LEAFs, e.g., with an unsupervised autoencoder[12,28,29,39,59] (Figure S4), and for integration with the supervised models utilising property-specific elemental descriptors in a variety of downstream tasks for materials property prediction. Such integration can be performed as follows:

$$\mathfrak{H}(\mathbf{a} \cdot \mathfrak{D}(\mathbf{M})) = y_a, \qquad (4)$$

where the base supervised model for property prediction $\mathfrak{H}$ acts on the input of a compositional vector $\mathbf{a}$ and the local environments matrix $\mathbf{M} = (m_{ij})^{21706 \times 86}$, connected via a dense layer $\mathfrak{D}(\mathbf{M}) = \sigma(\sum_{i,j}^{n} m_{ij} w_{ij} + b_i)$ with ReLU activation function[60] $\sigma$, kernel weights $w_{ij}$ and biases $b_i$, to predict a property $y_a$. For the considered example of classification of the Li-ion conducting materials ionic conductivity, integration of CrabNet with LEAFs as in Eq. (4) results in an equivalent average accuracy of 81% and an increased MCC of 0.60 over 5-fold cross-validation in comparison to the original results of CrabNet used with mat2vec.

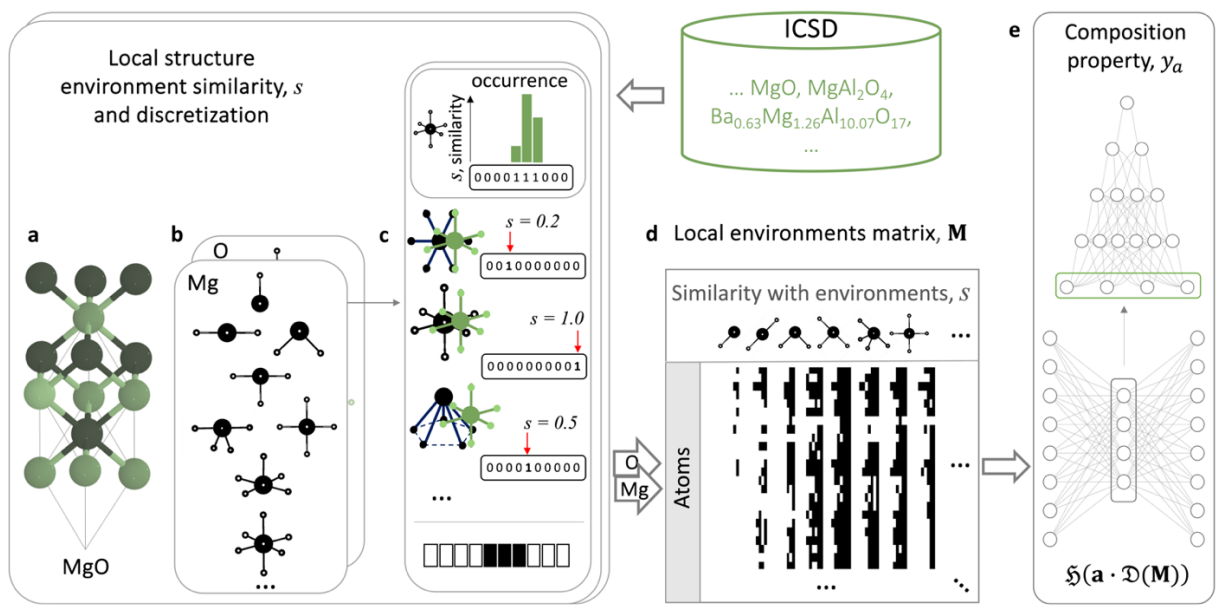

**Figure 4.** Schematic learning of local environment-induced atomic features (LEAFs) aligned with prediction of properties of materials represented solely by their compositions. Similarities of the local structure environments of the atomic sites in a crystal structure, exemplified by MgO (a), to the 37 selected common structural motifs[47] (b) are calculated for the atomic sites for experimentally verified structures reported in ICSD. (c) For the example of the six-coordinated Mg octahedral environment in MgO, compared to planar hexagonal (similarity, $s = 0.2$), octahedral ($s = 1$), and pentagon pyramidal ($s = 0.5$) motifs, these similarity values, $s$, are discretised into a thousand bins spanning from 0.001 to 1, illustrated as 10-digit binary strings for Mg example in (c) for simplicity. Such discretisation and subsequent concatenation of the binary strings for all 37 structural motifs form 37,000-long vectors of 0s and 1s, denoting the absence or presence of the degrees of similarity to the particular motifs in ICSD for considered chemical elements. The matrix of elemental local environments, **M**, is represented schematically as black (for ones) and white (for zeros) pixels in (d). Its dimensionality reduction by a single hidden layer neural-network autoencoder, $\mathfrak{D}$, to produce structure-induced elemental vectors, **a**, is trained end-to-end with a neural network, $\mathfrak{H}$, (e.g., CrabNet[54], plotted schematically with NN-SVG[61]) for prediction materials properties, $y_a$ (e).

Notably, such integration trained on the full dataset of 757 entries of conducting materials reported at all temperatures, achieves a higher accuracy of 77% and MCC of 0.53 in comparison to the 70% accuracy and MCC of 0.37 achieved with CrabNet with mat2vec, demonstrating enhanced robustness of the proposed approach to noise in the data arising from label ambiguity as the same compounds may have multiple conductivity entries at different temperatures. By employing the integration in Eq. 4 to train the models for other properties datasets such as dielectric, elasticity, formation energy, energy band gap, etc.[62], LEAFs demonstrate a comparable performance with the state-of-the-art models for compositions (Table 2), while offering a route for improved interpretability through connection to the prevalent structural features affecting the properties.

**Table 2.** Prediction of properties for composition-only description of materials: CrabNet with Mat2Vec vs CrabNet with LEAFs performance on MatBench datasets[62]

| Data set | Number of samples | CrabNet Mat2Vec | CrabNet LEAFs |
|---|---|---|---|
| | | \multicolumn{2}{c}{Mean Absolute Error} | |
| Perovskites form. energy (eV/unit cell) | 18928 | **0.3473** | 0.3495 |
| Dielectric (unitless) | 4764 | 0.4439 | **0.4254** |
| Elasticity G_VRH ($\log_{10}$(GPa)) | 10987 | 0.0994 | **0.0973** |
| Elasticity K_VRH ($\log_{10}$(GPa)) | 10987 | **0.0741** | 0.0761 |
| JARVIS exfoliation energy (meV/atom) | 636 | **49.8551** | 52.8234 |
| Experimental band gap (eV) | 4604 | 0.3463 | **0.343** |

## IV. CONCLUSION

LEAFs describe chemical elements in terms of the local structural motifs that are likely to form in crystalline inorganic solids. Learning atoms from crystal structures deepens our understanding of the chemical elements and their role in the composition-structure-property relationships. In practical terms, incorporating structural geometry into elemental descriptors enhances modelling of materials described solely as compositions and elucidates the role of particular atomic coordination geometries in determining materials properties. The LEAFs improvement over the state-of-the-art results is especially clear in tasks that are strongly correlated with structural information. In the structure-type classification, based on quantifying elemental similarity and likelihood of elemental substitution, LEAFs increase the state-of-the-art accuracy by 5% and improve the balance of multi-class assignment, as judged by MCC, by 0.09. This suggests the best practice for the popular materials design by substitution, where due to high-throughput approaches, a few per cent change in accuracy can result in thousands of new materials candidates. To facilitate this use of LEAFs, we provide an easy-to-use software tool[63] with simple commands for 1) measuring structure-induced similarity between materials (e.g., reported and hypothetical) described solely by their compositions, and 2) prediction of the most likely elemental substitution to retain structural stability for exploring novel materials. In contrast to other modern multi-dimensional descriptors, machine learnt from literature or materials data, which enable materials property modelling in many tests with accuracy comparable to LEAFs', LEAFs retain the direct links to structural motifs, enabling analysis of the elemental coordination environments underpinning composition-property relationships, e.g., through feature selection, thus making a step towards interpretable results of machine learning of materials. Complemented by the higher robustness to label noise for predicting material properties reported at different temperatures, as demonstrated on the example of Li-ion conductivity data, LEAFs will motivate integration of the proposed structural insights with other aspect elemental descriptors to further enhance materials modelling.

## Supporting Information

Details of the data processing[64], similarity calculations for local structure environments and their discretisation; autoencoder training; elemental similarity using LEAFs and LEAFs' distance for comparison of compositions; details of LEAFs' performance in crystal structure prediction – confusion matrices are available at https://www.github.com/lrcfmd/LEAF

**Data Availability Statement**

The data used in this study is available at https://www.github.com/lrcfmd/LEAF and available via the University of Liverpool data repository at https://doi.org/yyy/datacat.liverpool.ac.uk/xxx

**Acknowledgements**

We thank the UK Engineering and Physical Sciences Research Council (EPSRC) for funding through grant number EP/V026887.

**Keywords:** chemical element descriptor • local crystal structure • featurisation • machine learning of materials properties

# Supporting Information: Learning atoms from crystal structure

Andrij Vasylenko[1], Dmytro Antypov[1], Sven Schewe[2], Luke M. Daniels[1], John B. Claridge[1], Matthew S. Dyer[1], Matthew J. Rosseinsky[1,*]

[1]Department of Chemistry, University of Liverpool, Crown Street, L69 7ZD, United Kingdom

[2]Department of Computer Science, University of Liverpool, Ashton Building, L69 3DR, United Kingdom

*rossein@liverpool.ac.uk

Table of contents

Similarity calculations for local structure environments

In calculations of local structure environment of atomic sites we employ local structure order parameters(LoStOPs)[1] to quantify the agreement between a given observed coordination environment and the perfect elementary target environments in terms of angles. The elementary target motifs, such as 'linear', 'water-like', 'tetrahedral', etc. are illustrated in part in Figures 1 and 3 in the main text and in full in the original study[1]. In this approach, atomic site coordination is determined based on the Voronoi tessellation, and rescaling of the solid angle weights (defined by the Voronoi polyhedron) with the site properties, such as electronegativity differences and distance cut-offs. The resemblance between the local coordination environment of a given atomic site with a range of target motifs is calculated as maximum motif resemblance

$$q_a = \max(\{q_{a,j}\}) \qquad (S.1)$$

with individual motifs resemblance $q_{a,j}$ calculated with one single neighbor $j$ as the North pole for resemblance evaluation to motif type $a$ around the central site, their values vary smoothly between 0 and 1. For example, for the T-shaped coordination environment, $q_T$ LoStOP is calculated as

$$q_T = \max_{j \in N, k \neq j} \left\{ \sum_{l \neq j}^{N} \exp\left[-\frac{(\theta_{jl} - 90°)^2}{\Delta\theta^2}\right] \cos^2 \varphi_{jkl} \right\}, \qquad (S.2)$$

where $N$ is the number of nearest neighbours, $\theta_{jl}$ is an angle between the North pole neighbour $j$, central atomic site and neighbour $l$, $\Delta\theta$ is a parameter penalising angle difference with 90°, $\varphi_{jkl}$ is angle of a prime meridian (Figure S1).

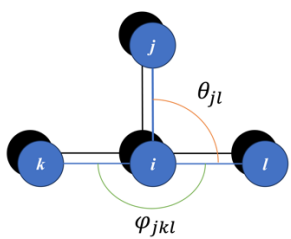

**Figure S1. Structural motif similarity on example of the T-shaped motifs (adapted from Ref.[1]).** Comparison of the motif for the central atom $i$ (blue circles) with the target T-shaped motif (black circles) in terms of the angles formed by the neighbouring atoms $\theta_{jl}$, and $\varphi_{jkl}$.

The resemblance values $q_a$ for all atomic sites with 37 target motifs are calculated with the LoStOPs implementations in Matminer[2] for 200809 inorganic crystal structures reported as Crystallographic Information Files (CIFs) in Inorganic Crystal Structure Database (ICSD)[3] (accessed 7.9.2021), which are processed with Pymatgen's CifParser[4]. The calculated values of structural motif similarities for this crystal structural data form the basis for LEAFs and extended matrix of local environments **M** presented in the main text. For example, for Mg-atom in MgO, the similarity values, $s$, to 37 target motifs are presented in Figure S2.

## Building LEAFs from similarities to local environments

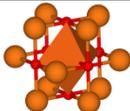
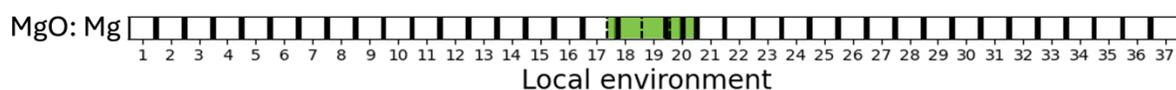

**Figure S2. Similarities of Mg local structure environment in MgO to common structural motifs and representation of chemical element (Mg) as a binary 370-long vector.** Among 37 considered motifs, 34 motifs are dissimilar with Mg local environment in MgO ($s$ = 0), and through discretization in 10-bin one-hot encoding, for illustration, can be represented as (1 0 0 0 0 0 0 0 0 0), where only the first bin denoting $s$ =0

contains a value. Six-coordinated motifs (highlighted with green), hexagonal planar ($s$ =0.2), pentagonal pyramidal ($s$ =0.5) and the most similar octahedral ($s$ =1) can be represented as (0 0 1 0 0 0 0 0 0 0), (0 0 0 0 1 0 0 0 0 0), and (0 0 0 0 0 0 0 0 0 1), respectively. Concatenation of these similarities forms a 370-vector for Mg local environment represented as white and black stripes (for 0s and 1s, accordingly). In this vector, the six-coordinated representations are divided with the dashed lines, the 0s are highlighted with green, and the 1s, represented with black, are located in the 2$^{nd}$, 10$^{th}$, and the 5$^{th}$ positions, respectively.

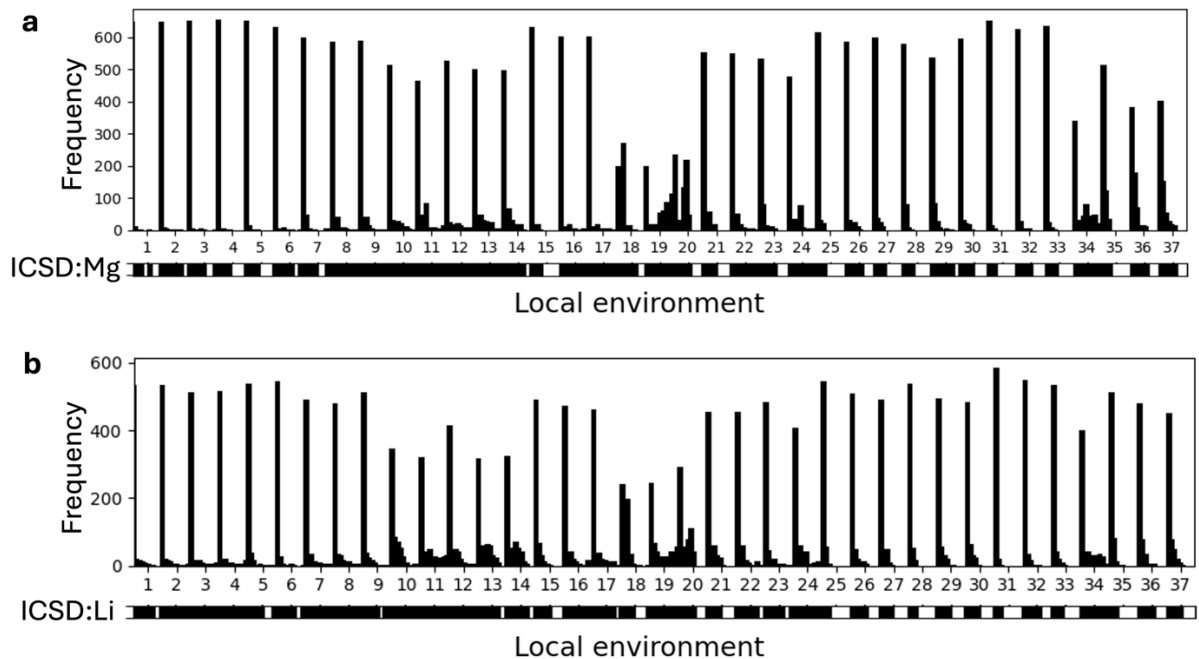

**Figure S3. Similarity values for Mg (a) and Li (b) to 37 common local structure motifs collected from all materials reported in ICSD, illustrated as 370-bin histograms and corresponding 370-bit binary strings below.** Visual similarity of the Mg and Li binary vectors, presented here with 370 bits for illustration, is further reduced by 1000-bit discretization of structural motifs resulting in 37000-bit elemental vectors, used in this work.

LEAFs autoencoder training

In the unsupervised setting, LEAFs can be calculated through dimensionality reduction of the matrix of local environments $\mathbf{M} = (m_{ij})^{21706 \times 86}$ to $(\widetilde{m}_{ij})^{n \times 86}$, where 86 signifies the number of elements, and 21706 is the number of non-zero columns resulting from discretization of the elemental 37 LEAFs values into 1000 bins each as described in the main text. Dimensionality reduction is achieved by training a single latent layer of size *n* shallow autoencoder neural network, while minimising the loss function – the reconstruction error, which in this context, is the Euclidean distance between the decoded output vectors and the original input vectors, which constitute $\mathbf{M}$ matrix. The best training results for *n* = 59 are achieved with the loss function calculated as the mean squared error as presented in Figure S4 and fixed learning rate 1e-5, demonstrating convergence of the loss on the held-out test data after 500 epochs.

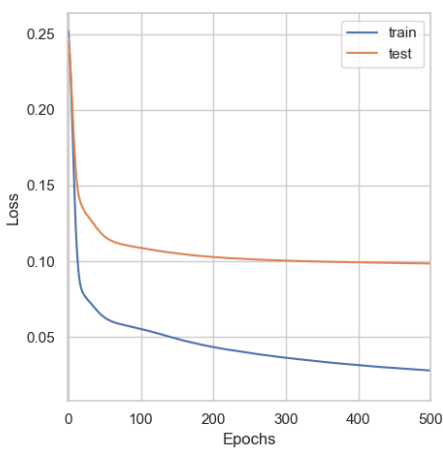

**Figure S4. Training of the LEAF shallow autoencoder with mean squared error loss.**

Elemental similarity

Chemical elements represented as vectors with LEAFs can be compared with cosine similarity. When arranged by increasing atomic number, the two-dimensional cosine map unveils similarity trends between the elements. Such maps can also qualitatively highlight differences in various elemental descriptors as proposed in Ref.[5] (Figure S5). High cosine similarity between chemical elements represented with LEAFs can serve a reliable basis for element substitution as defined in Eq. 1 in the main text, while retaining the structure of the host material (Table 1 in the main text and Figure S6). The partition function, *Z*, in Eq. 1, is calculated as a sum over all cosine similarity values in

Figure S5: $Z = \sum_{i,j} e^{\cos(\mathbf{a}_i, \mathbf{a}_j)}$, where indices $i, j$ label different elements in a pair, and summation is performed over all elemental pairs.

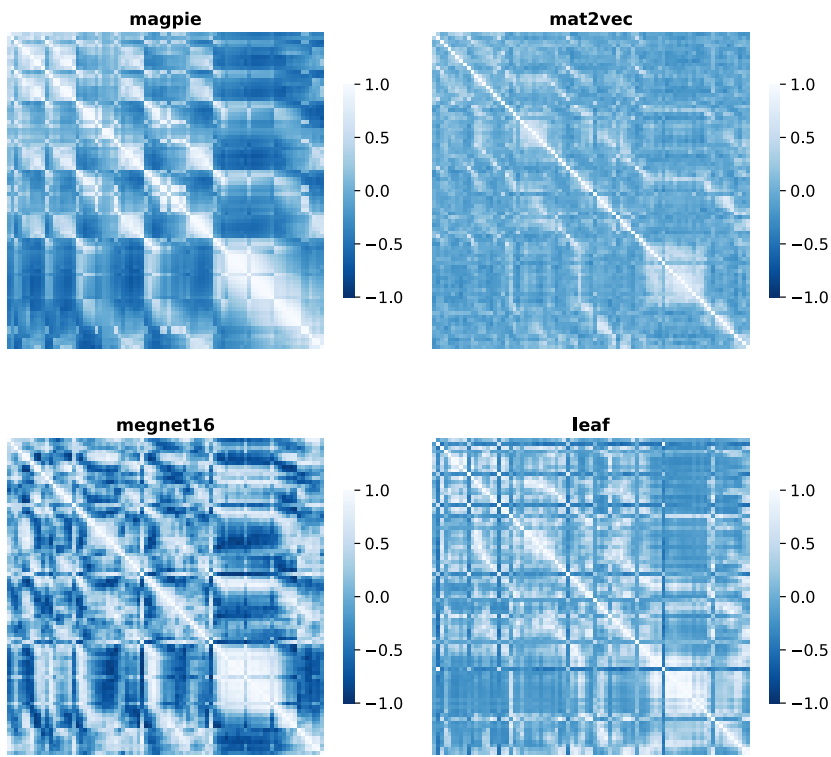

**Figure S5. Cosine similarity between chemical elements described with various elemental features (adapted from Ref [6]) highlights different grouping trends arising, depending on the features used.**

LEAFs' performance in crystal structure type multi-class classification

Accuracy and MCC performance metrics for this test presented in Table 1 in the main text can be derived from the detailed confusion matrices (Figure S6), in which on-diagonal elements depict the numbers of correctly classified structure types and off-diagonal numbers depict classification errors. LEAFs demonstrate the highest values for on-diagonal numbers and the smallest values for off-diagonal number, illustrating the best performance in classification.

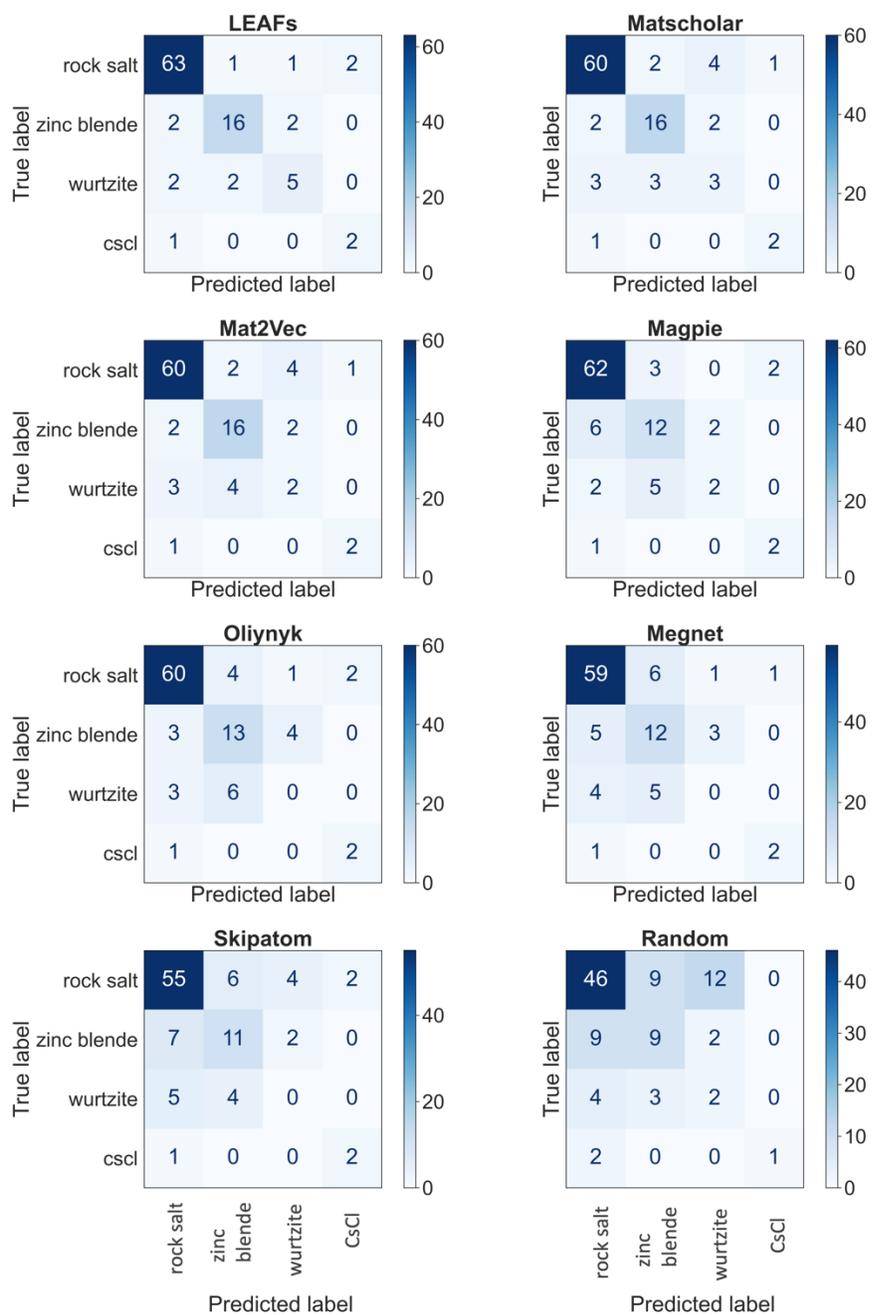

**Figure S6.** Confusion matrices obtained in classifying crystal structure types for compositions described with different elemental feature sets (adapted from Ref. [6]).

LEAFs distance for composition comparison

Chemical elements represented with LEAFs have been demonstrated to cluster according to chemical trends in Fig. 2a in the main text. In contrast, the analogous t-SNE map for chemical elements represented with random values vectors of the same size as LEAFs does not show any meaningful grouping in Fig. S7.

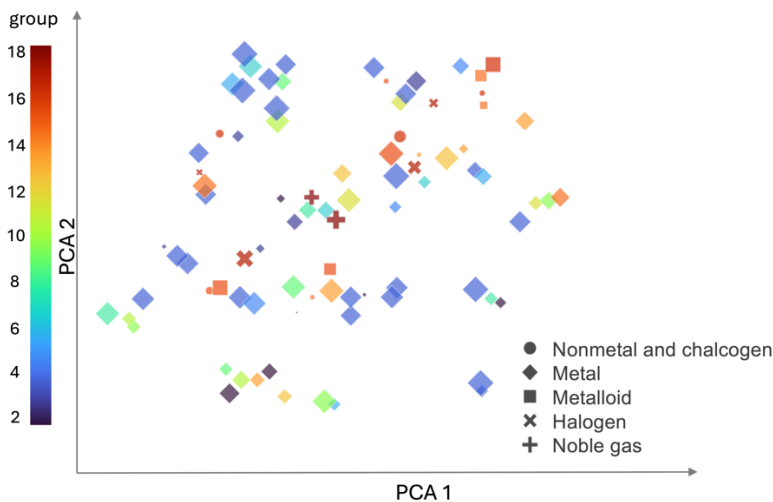

**Figure S7.** t-distributed Stochastic Neighbour Embedding (t-SNE) map of chemical elements represented with 37-bit vectors of random numbers contrasts with the emerging chemical trends in the analogous t-SNE map produced for elements represented with 37-bit LEAFs (Figure 3a in the main text).

Moreover, clustering of materials regarding their structure type observed in Fig. 2b in the main text and in Figure S8 indicates that distances in the LEAFs-represented chemical space capture structural relationships and can be used as a metric for mapping. For example, the metric for measuring similarity between the compositions can be expressed as cosine similarity between the corresponding compositional representations with LEAFs (Eq. 2 in the main text):

$$S(\mathbf{a}_{Li_3PO_4}, \mathbf{a}_{Li_7La_3Zr_2O_{12}}) = e^{\cos(\mathbf{a}_{Li_3PO_4}, \mathbf{a}_{Li_7La_3Zr_2O_{12}})}, \qquad (S.3)$$

with the exponent magnifying the differences in the multi-dimensional space.

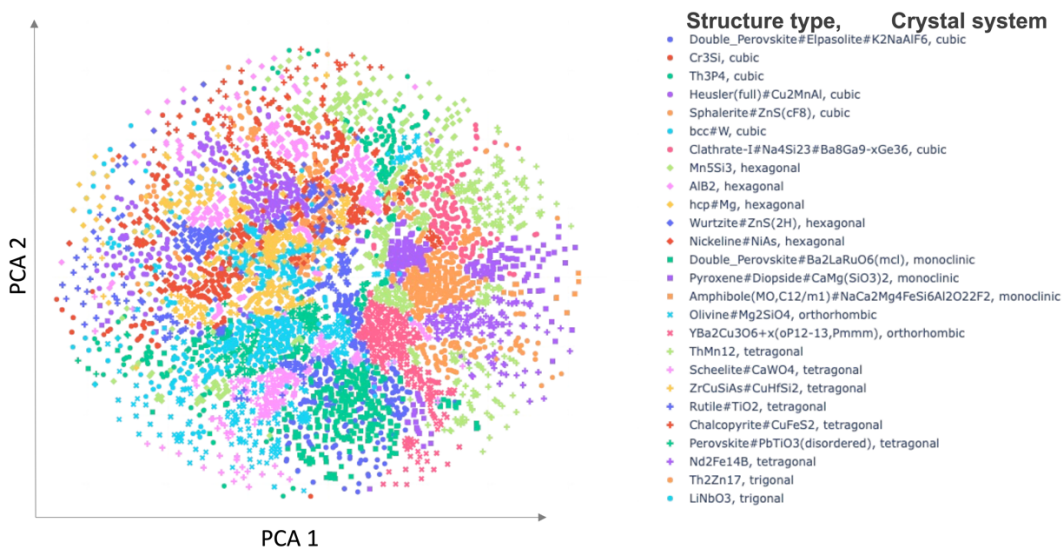

**Figure S8. t-SNE map of the LEAFs-represented compositions forming structure types represented with $N \in [500, 1000]$ compositions with original notations in ICSD[3].** With varying compactness of distribution in two principal dimensions, all clusters are located in a distinct area of the map, correlated with the structure types and crystal systems they represent.

LEAFs' utility as elemental representation in composition-property ML models

Digital representation of chemical elements is essential for materials modelling. In the small data regimes that are prevalent in materials science, the choice of representation can have a significant impact on the model performance, especially for the models relying on composition-only input. We demonstrate the utility of LEAFs for such models, trained with CrabNet[7] in integration with the local environments matrix as described in Eq. 4 in the main text. The parameters of CrabNet for the training on the representative materials datasets[8] are unchanged from the original CrabNet study, where Mat2Vec[9] elemental features were employed instead of LEAFs. The two approaches are compared in terms of the average Mean Absolute Error (MAE) computed for the 5-fold cross validation (Table 2 in the main text), in which Mat2Vec and LEAFs representations demonstrate overall comparable accuracy in six tests, with maximum MAE improvement of 4% for LEAFs in Dielectric test, and 6% for Mat2Vec in JARVIS exfoliation energy test.

In contrast to engineered or machine learnt digital elemental representations, including Mat2Vec and random features, LEAFs enable analysis of the elemental local structural environments underpinning composition-property relationships, e.g., through feature selection (Figure 3a in the main text). Additionally, the structures of the Li-ion conducting materials[10] can be analysed in terms of the features, rendered important in Figure 3; in Figure S9, the distribution of such top 25 local structural environments for lithium atomic sites corresponding to high Li-ion conductivity in reported compounds[10] is illustrated.

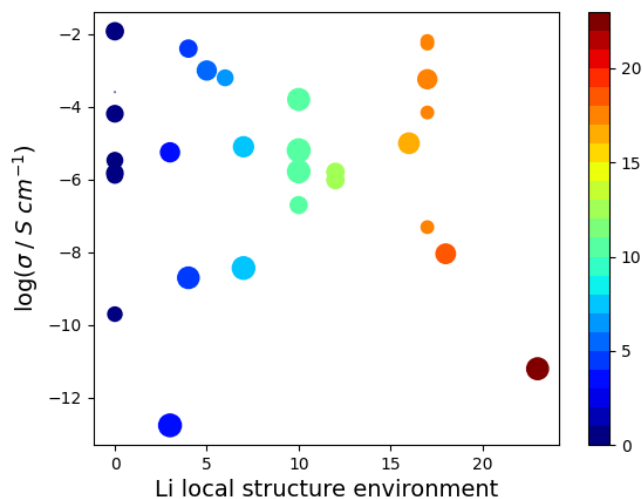

**Figure S9. The most pronounced Li local structure environent in Li-ion conducting materials and their respective conductivity.** The x-axis sequentially enumerates the considered Li local structure environments (also colour-coded) listed in full in Figure S2 , with the first 10 labelled in Figure 3 in the main text. The size of the markers corresponds to the degree of similarity of the Li coordination to the corresponding structural motif. The broad distribution of local structural environments for Li sites corresponding to high Li-ion conductivity in reported compounds[10] indicates the absence of a specific preferred Li coordination associated with high Li-ion conductivity.